\documentclass[a4,12pt]{article}
\usepackage{amsmath,amsfonts,amssymb,graphicx,epsfig}
\usepackage{colordvi}
\def\ba{\begin{eqnarray}}
\def\ea{\end{eqnarray}}

\begin{document}
\title{\textbf{Analytical and Monte Carlo study of two antidots in magnetic nanodisks with vortex-like
magnetization}}
\author{A.R. Pereira\\ \small\it Departamento de F\'{\i}sica, Universidade Federal
de Vi\c{c}osa,\\ \small\it 36570-000, Vi\c{c}osa, Minas
Gerais, Brazil \\ \small \it Physics Department,
University of Bologna, Via Irnerio 46, I-41126, Bologna, Italy.\\
A.R. Moura and W.A. Moura-Melo\\ \small\it Departamento de F\'{\i}sica, Universidade Federal
de Vi\c{c}osa,\\ \small \it 36570-000, Vi\c{c}osa, Minas
Gerais, Brazil\\
D.F. Carneiro, S.A. Leonel, and P.Z. Coura \\ \small \it
Departamento de F\'{\i}sica, ICE, Universidade Federal de Juiz de
Fora,\\ \small \it 36036-330, Juiz de Fora, Minas Gerais, Brazil}

 \date{}
\maketitle
\begin{abstract}
How stable vortex-like magnetization in magnetic nanodisks with
small aspect ratio ($L/R<<1$) is affected by two antidots is
investigated analytically and by Monte Carlo simulations. For
suitable ranges of the physical parameters this vortex presents
bistable states when pinned around the antidots. The hysteresis
loop obtained shows a central loop associated to the pinning
mechanism of the vortex at the antidots. Our results agree
qualitatively well with those provided by experiments and
micromagnetic simulations.
\end{abstract}

\newpage
\indent Recently, ferromagnetic disk-shaped nanostructures with
submicrometer lateral dimension (thickness) $L<1\mu{m}$ have been
fabricated and investigated for their potential applications in a
number of  magnetoelectronic mechanisms. In particular, it has
been observed that above the so-called single-domain limit,
magnetic vortex states appear in these samples, exhibiting a
planar-like arrangement of spins outside the core, where a
perpendicular magnetization is observed\cite{vortex-observed-exp}.
As long as one could manipulate these states other possibilities
would emerge. In fact, one way towards this control is obtained by
removing some small portions of the magnetic nanodisk, in such a
way that the cavities (antidots) so created work by attracting and
eventually pinning the vortex around themselves \cite{RahmPRL95,RahmAPL82,RahmJAP95,AfranioPRB2005,AfranioJAP2005,LEO}.
Based upon such an idea, Rahm and coworkers \cite{RahmAPL85} have studied the cases of two, three and four antidots (each
of them with diameter $\sim85\,{\rm nm}$) inserted in a disk with
diameter $\sim 500\,{\rm nm}$, separated by around $150\,{\rm
nm}\,-\,200\,{\rm nm}$. Their experimental results confirmed the
previous statement about vortex pinning and put forward the possibility
of using these stable states as serious candidates for magnetic memory and
logical applications as long as we could control vortex position,
for example, applying a suitable external magnetic field which
should shift the vortex core from one defect to another, and
vice-versa. Basic logical operations have been obtained by means
of bistable magnetic switching\cite{RahmAPL87}.
Although experimental results are provided for this system a suitable
analytical analysis is still lacking. The latter is important, for
instance, to provide the basic physics behind the mechanism of vortex pinning and
switching processes, giving the relevant parameters for a better
control of possible applications and also indicating the
limitations. Our present analytical model and Monte Carlo (MC)
simulations have been able of capturing the basics involved in
this problem, namely, how the vortex experiences the effects of
the defects and how hysteresis loops are sensitive to the latter.

Let us starting by considering a magnetic dot represented by a
small cylinder of radius $R$ and thickness $L$ (so that its aspect
ratio $L/R <<1$). In addition, we shall assume that along the
axial direction ($z$-axis), the magnetization $\vec{M}$ is
uniform. Furthermore, if we introduce $N$ isolated holes (each of
them with height $L$ and radius $\rho<<R$) in the dot, the total
magnetic energy of the nanodisk can be approximated, in the
continuum limit, by  \ba E_{\rm mag}=\frac{L}{2}\int\int_D
\left[A(\partial_\mu\vec{m})\cdot(\partial^\mu\vec{m}) -M^2_s\,
\vec{m}\cdot(\vec{h}_m +2\vec{h}_{\rm ext})\right]
\prod^N_{i=1}\,U_i(\vec{r}-\vec{d}_i)\, d^2r \,, \ea where $A$ is
the exchange coupling, $D$ is the area of the cylinder face,
$\vec{m}=\vec{M}/M_s$ is an unity vector describing magnetization
along $D$ (with $M_s$ being the saturation magnetization),
$\vec{h}_m=\vec{h}_m(\vec{m})\equiv \vec{H}_m/M_s$ is the
demagnetizing field, $\vec{h}_{\rm ext}$ is an applied magnetic
field (Zeeman term) and $\mu=1,2$. The potential $U$, in turn,
brings about the effect of the antidots distributed throughout the
nanodisk, say,
$\prod^N_{i=1}\,U_i(\vec{r}-\vec{d}_i)=U_1(\vec{r}-\vec{d}_1)\,U_2(\vec{r}-\vec{d}_2)\ldots
U_N(\vec{r}-\vec{d}_N)$, with \ba U_i(\vec{r}-\vec{d}_i)=
\left\{\begin{array}l 0 \quad{\mbox{ \rm if}}\quad
\mid \vec{r}-\vec{d}_i \mid <\rho\\
                              1 \,\quad{\mbox{ \rm if}} \quad \mid \vec{r}-\vec{d}_i
                                 \mid \geq \rho  \end{array}\right.\,. \label{Uimp}
\ea Therefore, the system of a dot with $N$ isolated antidots may
be viewed as a cylinder of radius $R$ and thickness $L$ with $N$
smaller cylindrical cavities with radius $\rho<<R$, each of them
centralized at $\vec{d}_i$. Here, we shall study explicitly the
case $N=2$ (the treatment for $N>2$ may be performed in the same
way). Experimental and numerical results are available for very
similar systems, say, disks with $L\sim \,30\,{\rm nm}$, $R\sim
\,500\,{\rm nm}$ whose antidots (up to four) have diameters
$2\rho\sim\, 80\,{\rm nm}$ separated by $150\,{\rm nm}\,-\,
200\,{\rm nm}$ \cite{RahmAPL85}.

Now, let us consider a cylindrically symmetric vortex-like
magnetization throughout a dot, say, with the vortex core
centralized at $\vec{r}=\vec{0}$. This is the ground state of a
nanodisk in the absence of holes and external magnetic fields. For
that, it is convenient to write $\vec{m}=(\sin\theta\cos\varphi,
\,\sin\theta\sin \varphi, \, \cos\theta)$, with
$\theta=\theta_v(r)$ and $\varphi=\arctan(y/x)\pm\pi/2$. The
function $\theta_v(r)$ may be approximated by $\sin\theta_v(r)=0$
in the dot center ($\vec{r}=\vec{0}$), while $\sin\theta_v(r)\to1$
far away the center, $\mid\vec{r}\mid=r>>a$ ($a=\sqrt{A}/M_{s}$ is
the unit-cell element size or the exchange
length; for most magnetic materials $a=5-6 \,{\rm nm}$). In words, the magnetization consists of a small
core where spins display out-of-plane components for regularizing
the exchange energy, and an outer region where spins are
practically confined to the dot plane face. In this case, the
magnetic superficial charges in the lateral face of the dot and
the magnetic volumetric charges ($\vec{\nabla}\cdot\vec{m}$)
identically vanish yielding no contribution to $\vec{h}_m$. The
antidots affect this picture as follows: from the point of view of
the exchange term, antidots lead to a less exchange energy for the
vortex. Then, the topological structure is attracted by the hole
suffering modifications in its profile. In addition, the
distribution of magnetic charges throughout the internal edges of
the cavities (holes) and mainly along the external lateral face of
the cylinder (whenever the antidot is not at the geometrical
center of the dot) increase the magnetostatic energy due to a
change in the product $\vec{m}\cdot\hat{n}_{s}$ ($\hat{n}_{s}$ are
unit vectors normal to external lateral surface of the disk and
internal surfaces of the cavities). Thus, the demagnetizing field,
$\vec{h}_{m}$, can be obtained from its associated potential
$\Phi_m=\Phi_{\rm V}+\Phi^e_{\rm edge}+\Phi^i_{\rm edge}$, in the
usual way, $\vec{h}_m=-\vec{\nabla}\Phi_m$. Here, $\Phi_{\rm V}$
is the magnetostatic potential related to the volumetric charges ,
while $\Phi^e_{\rm edge}$ and $\Phi^i_{\rm edge}$ comes about from
the surface charges on the external and internal (holes) edges,
respectively. The contributions of the volumetric potential can be
neglected since the approximations considered above leads to
$\vec{\nabla}\cdot\vec{m}=0$.

\begin{figure}[h!]
\centering \hskip 1cm 
\fbox{\includegraphics[width=8cm,height=3cm]{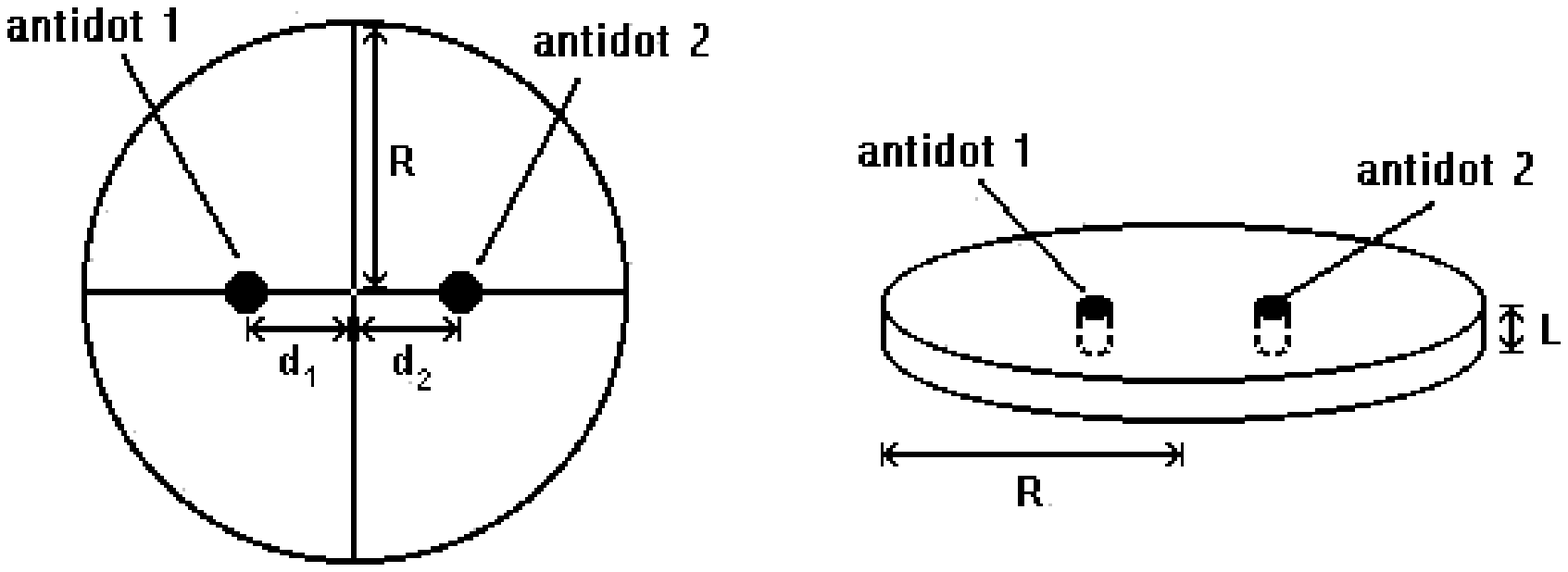}} \vskip
-0.0cm \caption{} \label{Figure1}
\end{figure}

For simplicity, the antidots centers are assumed to lie along a
straight line that crosses the origin of the dot, say, at
$\vec{d}_1$ and $\vec{d}_2$, as shown in Figure \ref{Figure1}. We
also assume that the vortex displacement $\vec{l}$ from its
equilibrium position is not too large, so that it experiences no
appreciable change in its profile (`rigid' vortex behavior
\cite{Guslienko}). Thus, the exchange potential experienced by the
vortex may be estimated as \cite{FagnerPLA2004} \ba V_{\rm
ex}(s,d_{1},d_{2})\cong
\frac{\pi\,A}{2}\,\ln\left[(1-s^2)\left(1-f_1-f_2+f_1f_2\right)\right]\,,
\label{Vex} \ea where $\vec{s}=\vec{l}/R$, supposed to be small
($\mid \vec{s}\mid <<1$), measures the relative shift of the
vortex from the dot center; the functions $f_{i}$ are defined as
$f_i=a^2/(\mid \vec{d}_i-\vec{s}R \mid^2+b^2)$, where $b\cong
1.147\rho$ is a constant introduced to avoid spurious divergences
whenever the vortex is centralized at one of the defects
\cite{AfranioPRB2005,LEO}. Note that each $f_i$ is related to the
attractive potential that each isolated hole induces on the
vortex, while the product $f_1\,f_2$ accounts for a competition
between them \cite{FagnerPLA2004}. Thus, if our system were large
enough, the vortex would displace towards one of the holes as
shown in Ref.\cite{FagnerPLA2004}. In a small magnet, things are
much more interesting once the vortex experiences a modification
in its profile and magnetization is no longer cylindrically
symmetric. Then, the magnetostatic energy increases and a
restoring force appears in order to pull the vortex back to the
dot center. Indeed, for small displacements of the vortex we may
estimate this energy shift analytically, like below
\cite{AfranioPRB2005,Guslienko}
\begin{eqnarray}\label{Vmag}
V_{mag}(s,\vec{d}_1,\vec{d}_2,\rho)\cong & 2\pi^2 M^2_s
(R^2-2\rho^2) \big\{F_1(\frac{L}{R})s^2 +
F_1(\frac{\rho}{R})\big[\alpha
\big(s-\frac{\mid \vec{d}_1\mid}{R}\big)^2 &  \nonumber \\
&+\beta \big(s-\frac{\mid \vec{d}_2 \mid}{R}\big)^2 + \gamma
\left(\frac{\rho}{R}\right)^2\big] \big\},&
\end{eqnarray} \\
in which the first term is related to the contribution of the disk
envelop, while the remaining ones are associated to the pinning of
the vortex by one of the defects. In addition,
$F_1(\xi)=\int^\infty_0
\,\frac{J^2_1(t)}{t}\left(1-\frac{1-e^{-\xi{t}}}{\xi{t}}\right)\,dt$,
where $J_1(t)$ is the Bessel function. The possible three distinct situations are: i) the vortex
is centered at the antidot 1, then $\alpha=1$ and
$\beta=\gamma=0$; ii) it is centered at the antidot 2, thus
$\alpha=\gamma=0$ while $\beta=1$; iii) the vortex is not pinned
at any defect, what yields $\alpha=\beta=0$ and $\gamma=1$.

\begin{figure}[h!]
\centering \hskip 1cm 
{\includegraphics[width=8cm,height=12cm]{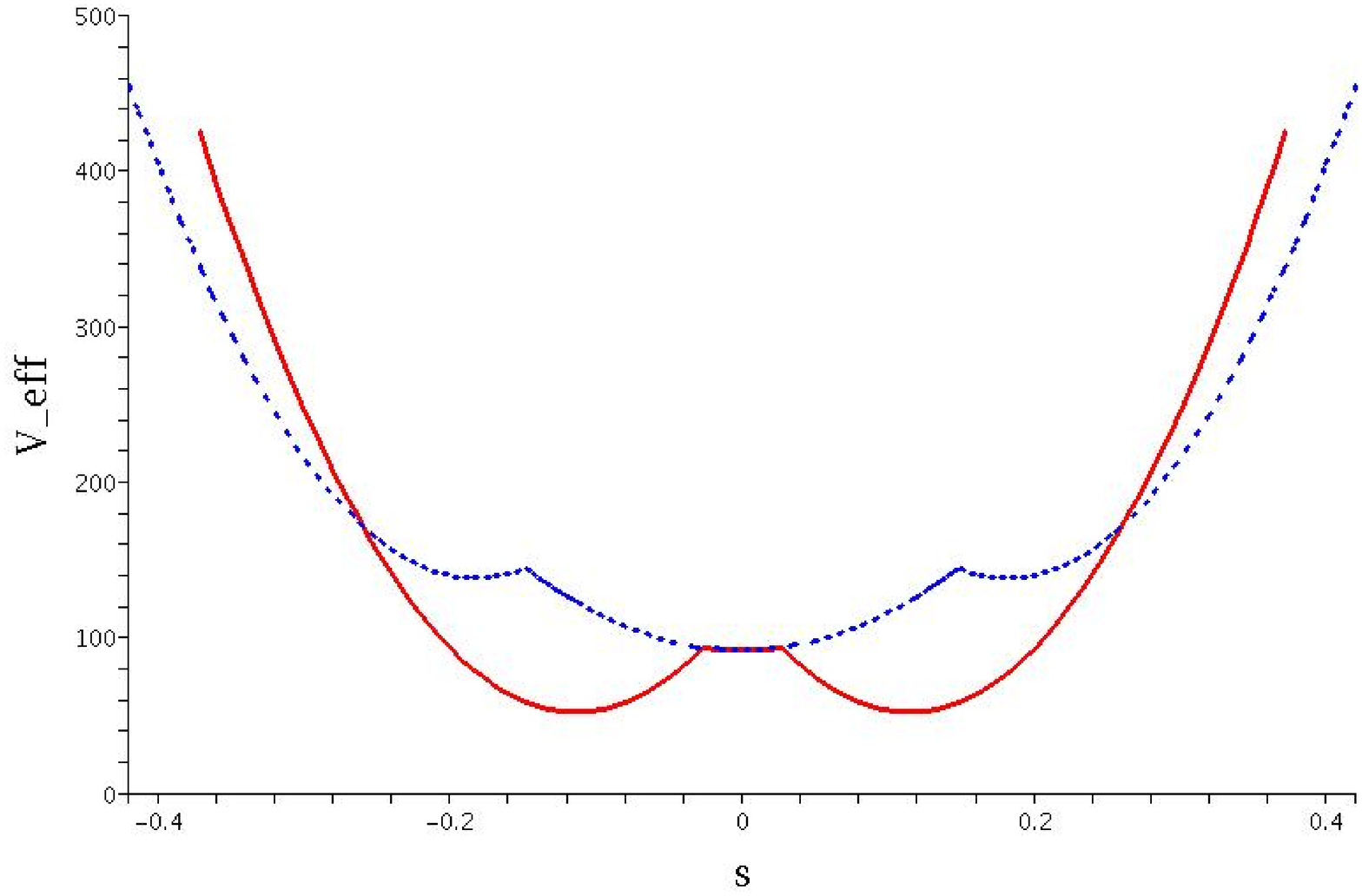}}
\caption{} \label{Figure2}
\end{figure}

Figure \ref{Figure2} shows how the effective potential, $V_{\rm eff}=V_{\rm ex}+V_{\rm mag}$,
behaves as function of $s$. Bistable states corresponding to
the vortex-antidot pinned configurations appear for suitable
values of the parameters. Our analytical model predicts that three
parameters are the most relevant in this case: the radius of each
antidot $\rho$, their center-to-center separation,
$D\equiv|\vec{d}_1-\vec{d}_2|$, and the (characteristic) exchange
length $a=\sqrt{A}/M_s$. We have observed that the appearance of
bistable states in $V_{\rm eff}$ are generally associated to
larger values for $\rho$ and $a$ and to smaller $D$, as
illustrated in Fig. \ref{Figure2}. Actually, it shows that our
analytical results qualitatively agree with those obtained in
experiments of Ref.\cite{RahmAPL85}, but there the bistable states
are observed even for larger separations ($\sim 200\,{\rm nm}$)
than those fitted by the present analysis ($\sim 130\,{\rm nm}$).
This discrepancy is related to the small displacements of the
vortex core assumed by our model.

In the presence of the two holes the vortex equilibrium
position, $s_0=(x_0,y_0)$, can be easily determined by evaluating
$dV_{\rm eff}/ds=0$. Clearly, depending on the relevant
parameters, the geometry of our system implies that $s_0$ will
always correspond to ($\mid x_0\mid \geq 0, \,y_0=0$) so that the
general expression for the local magnetization along the dot face
now reads:
\ba m_x(x,y,s_0)=\frac{\mp
y}{\sqrt{(x-s_0R)^2+y^2}}\,,\qquad m_y(x,y,s_0)=\frac{\pm (x-s_0
R)}{\sqrt{(x-s_0R)^2+y^2}}\,,
\label{mxmy} 
\ea
where the upper (down) signs are associated to counterclockwise (clockwise)
magnetizations. Since the system is antisymmetric under reflection
against $x$-axis, the average magnetization along this axis
vanishes, while the $y$ component may be estimated like below (the factor
$2$ in the 2nd term accounts for 2 antidots):
\ba
\left< M_y\right> \cong M_s\left[\frac{1}{\pi\,R^2}\int _{\rm
disk} m_y(x,y,s_0)\,dxdy -\frac{2}{\pi\,\rho^2}\int_{\rm antidots}
m_y(x,y,s_0)\,dxdy\right]\,.
\label{Mymedia}
\ea

In addition, if an external homogeneous magnetic field is applied
along the $y$-axis, $\vec{h}_{\rm
ext}=\vec{H}/|\vec{M}_s|=h_y\hat{y}$, then, in the lowest order,
$V_{\rm eff}$ must be augmented by : \ba V_{\rm h_{\rm ext}}(s)=
-\pi M^2_s h_y\,(R^2-\rho^2)\,(s-s_0)\,,\label{VZeeman} \ea where
$s$ is the shift in the vortex center position caused by the
field. So the total potential now reads $V_{\rm total}=V_{\rm eff}
+V_{\rm h_{\rm ext}}$. In the presence of $h_{\rm ext}$
the vortex equilibrium position is now $s_h\neq\,s_0$.
For example, if one takes the counterclockwise magnetization
(upper signs in eq. (\ref{mxmy})), then for $h_y>0$ ($<0$) the
vortex will be shifted to the left (right) of $s_{0}$. Thus, the
equilibrium position of the vortex center in the presence of
the applied external field, $s_h$, is simply a small shift of
$s_0$ along the $x$-axis. Therefore, in this situation
$\left<M_x\right>=0$ while $\left<M_y\right>$ is calculated from
eq. (\ref{Mymedia}) with $m_y(x,y,s_h)$ instead of $m_y(x,y,s_0)$.
How magnetization behaves as $h_{\rm ext}$ is varied gives the
hysteresis loop left by the vortex motion under such conditions.\\

\begin{figure}[h!]
\centering 
{\includegraphics[width=8cm,height=15cm]{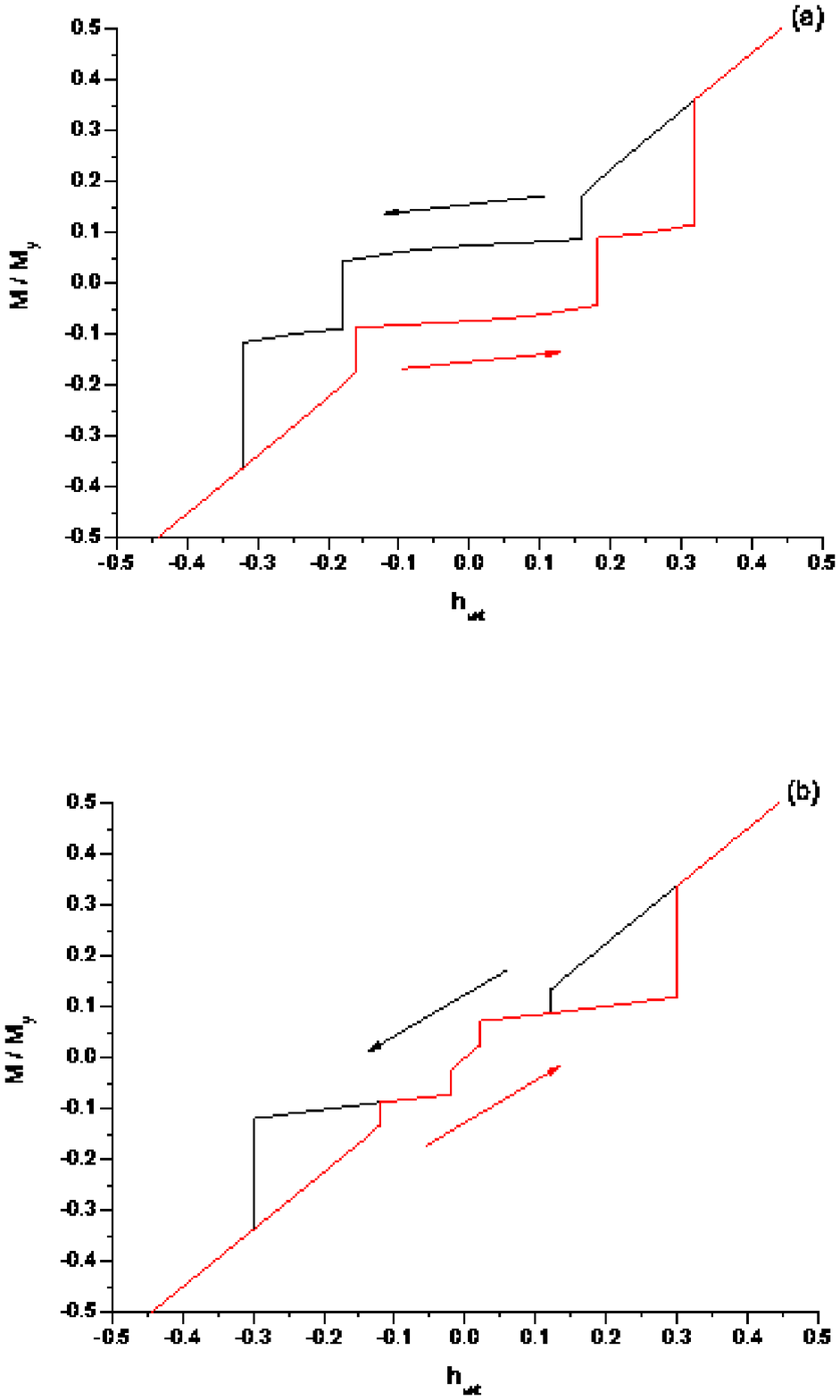}} \caption{} \label{Figure3}
\end{figure}

Figure \ref{Figure3} displays the hysteresis loops for
some values of the relevant parameters when an alternating-like
magnetic field is applied along the $y$-axis. The central loop
shown in this figure, on the left, is in qualitative agreement
with the experimental results of Refs.\cite{RahmAPL85,RahmAPL87}.
In these works, such a mechanism was observed for larger cavities
radii and separations ($\rho\sim 40\,{\rm nm}$ and $D\sim
200\,{\rm nm}$). In our analysis, we could observe a similar fact
only for much smaller values of these parameters. Again, such a
discrepancy comes about once our model is strictly valid for small
$s$. The central loop, taking place at magnetic fields weaker than
that for vortex annihilation, manifests the most important physical
feature of the system: the existence of two metastable vortex states
with the equilibrium position of the vortex center pinned at each antidot.
There, the jumps are due to vortex core switching from one state to another
while the plateaux corresponds to pinned vortex states. The
mechanism of switching can be explained in the following way: in
the situation shown in Fig. \ref{Figure2}(a) the vortex
core can be initially pinned in equilibrium at one of the
antidots, for instance, at the left antidot. However, this
configuration may be perturbed if a strong enough external
magnetic field is applied, say, along the $y$-axis, so that the
vortex core is shifted to the another antidot. Inverting the
magnetic field the vortex comes back to the initial configuration,
but leaving a hysteresis loop in the complete round.
The plateaux observed in the central loop of experimental data has
a pronounced inclination respective to the magnetization axis,
which may be associated to the deformation of the vortex profile
on the dot face. Consequently, our preceding approach could not
fit this fact. At the attempt of understanding this mechanism, we have
performed MC simulations in order to improve the range of validity
of our former results. For that, we consider the xy-model on a
finite flat disc (of radius $R$) supplemented by a strong enough
surface border anisotropy. Although it cannot be considered as the actual magnetostatic
energy we expect that it could imitate its role, say, increasing the total
energy as long as the spins develop normal components at the borders of the
nanodisk and of the holes (2nd and 3rd terms below; see Ref.\cite{IvanovPRB68})).
Indeed, this anisotropy has the effect of keeping the original cylindric-like profile
of the vortex near these borders although allowing it to deform in other regions
along the dot face. Therefore, Hamiltonian reads:

\ba \label{discModel} H = -J\sum_{i,j}(S^{x}_{i}S^{x}_{j} +
S^{y}_{i}S^{y}_{j}) + B\sum_{k\in disk border}(\vec
S_{k}.\hat{m}_{k})^{2} + B\sum_{h\in holes border}(\vec
S_{h}.\hat{m}_{h})^{2}\,, \ea where $J>0$ is the exchange
ferromagnetic integral, $\vec
S_{i}=(S^{x}_{i},S^{y}_{i},S^{z}_{i})$ are the classical spin
vectors specified at the lattice sites $i$ and the first summation
is over nearest-neighbor spins. The other term represents the
surface anisotropy where the sum over $k$ includes only the border
sites of the disk and sum over $h$ includes only the border sites of the holes.
Consequently, the unit vectors $\hat{m}_{k}$ are perpendicular to the circumference
envelop of the disk and the unit vectors $\hat{m}_{h}$ are perpendicular to the
circumference border of the holes.  There, $B$ is a constant of single-ion surface
border anisotropy so that for $B>0$ ($<0$) the border spins
$\vec{S}_{k}$ and $\vec{S}_{h}$ tend to lie perpendicular (parallel) to
$\hat{m}_{k}$ and $\hat{m}_{h}$ direction respectively. Of course, here we chose the
case $B>0$. Among other features, such a term
enables vortex deformation on the dot face whereas keep its
configuration at the border (for a fixed $B$), so that we can now fit the inclination
of the central loop discussed above.\\

\begin{figure}[h!]
\centering \hskip 1cm 
\fbox{\includegraphics[width=7cm,height=5cm]{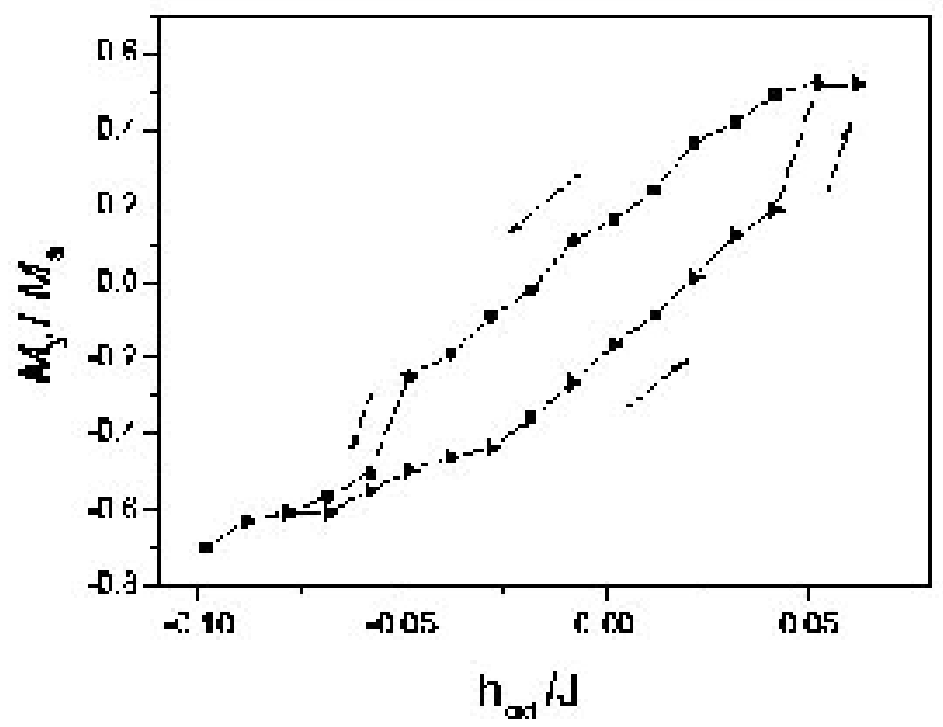}}
\caption{} \label{Figure4}
\end{figure}

In our MC simulations for hysteresis loop we have adopted Metropolis algorithm
\cite{Metropolis} with the initial configuration of centered
vortex in disks with diameters $2R=40a_{0}$ ($a_{0}$ is the
distance between two spins in a discrete lattice inside the disk).
In addition, we have introduced two circular holes (regions with
vacancy of spins) where we chosen radius $\rho=2a_{0}$ and center-to-center
separation $D=10a_{0}$, in such a way that $\rho/D\sim 0.2$ is close to that
from experiments (see Refs.\cite{RahmAPL85,RahmAPL87}). Besides, we
have taken the temperature ${ T/J = 0.05}$ and the parameter $B/J= 0.03$.
This is the critical value for $B/J$ so that, above it, non-centered vortices
are energetically favorable in these disks (see Ref. \cite{IvanovPRB68}).
Figure \ref{Figure4} shows a typical central hysteresis loop as the vortex
center is switched from one antidot to another and back again (note particularly
that its inclination is in good
qualitative agreement with experiments of Ref.\cite{RahmAPL85}).

In conclusion, our analytical as well as MC calculations have
shown how two holes incorporated into the body of a magnetic
nanodisk attract the remanent vortex-like magnetization to their
centers, creating the possibility of bistable states of
vortex-hole pinned configurations, as observed in experiments. The
analytical model for the effective potential (exchange +
magnetostatic) have yielded to this picture in good agreement with
experimental findings. However, concerning the hysteresis loops,
the analytical calculations show only a qualitative concordance
with experiments. Indeed, the appearance of a central loop in
those curves, related to the switching of the vortex core between
the two stable states is verified here for values of $\rho$ and
$D$ very smaller than those considered in experiments. In other
words, our results in this case would lead to a good agreement
with experiments only for very small displacements of the vortex
core around its equilibrium position, $s_h$ (`rigid' vortex
regime). It implies in small distance $D$ between the holes in a
nanodisk. For improving this scenario (considering as much as possible generic distances) we have also performed Monte Carlo
simulations which take into account the deformation of the vortex
and effects due to the border. Our simulations for the hysteresis
loop are in agreement to the experimental results and exhibits
bistable magnetic switching mechanisms. The approach developed
here also predicts the possibility of gyration of the vortex core
about one of the three equilibrium positions (see Fig.
\ref{Figure2}) with characteristic frequencies that depend on the
relevant parameters $R$, $\rho$ and $D$. The theoretical study of
a quite recent observed gyrotropic frequency of magnetic vortex
around antidots in nanosized structures\cite{Compton} is under
investigation and will be communicated elsewhere.

\vskip 1cm
\centerline{\large\bf Acknowledgements} \vskip .5cm
The authors thank CAPES, CNPq, and FAPEMIG (Brazilian agencies) for the financial supports.

\newpage

\section{Figure Captions}
Figure 1: Top (left) and lateral (right) views of a nanodisk with
radius $R$, containing two antidots lying along the $x$-axis,
whose centers are $\vec{d}_1$ and $\vec{d}_2$ apart from the disk center.\\

Figure 2: Typical plots of $V_{\rm eff}/A$ as a function of $s$.
Here, we have taken $R=250\,{\rm nm}$, $L=30 \,{\rm nm}$, $\rho= 43\,{\rm nm}$, and $a=\sqrt{A}/M_s=5.7 {\rm nm}$ (values considered in Ref.
\cite{RahmAPL87}). The defects are placed along
$x$-axis at the positions: (solid curve) $\vec{d}_1=\vec{d}_2=-50\,{\rm nm}$; (dashed curve) $\vec{d}_1=-\vec{d}_2=-80\,{\rm nm}$. Note that the bistable states tend to disappear as long as the distance between the cavities becomes larger.\\

Figure 3: The central loop associated to the switching of the vortex between the two antidots (bistable states) is jeopardized as long as $D$ increases and/or $a$ gets lower. External loops (not depicted) were associated to the creation/annihilattion of the vortex.   (a) $|\vec{d}_1-\vec{d}_2|\equiv{D}=40 \,{\rm nm}$, $\rho=10\,{\rm nm}$, and $a=\sqrt{A}/M_s=17 \,{\rm nm}$; (b) ${D}=40 \,{\rm nm}$, $\rho=10\,{\rm nm}$, and $a=5.7 \,{\rm nm}$.\\

Figure 4: Hysteresis loop for a disk with two antidots with radius
$\rho = 2 a_{0}$ and center-to-center separation $D = 10 a_{0}$
obtained by Monte Carlo simulations.
\newpage
\section{Figures}

\end{document}